\documentclass[twocolumn,showpacs,preprintnumbers,amsmath,amssymb]{revtex4}
\usepackage{tabularx,graphicx}\begin{document}
\newcommand{\beq}{\begin{equation}}
\newcommand{\eeq}{\end{equation}}
\newcommand{\beqn}{\begin{eqnarray}}
\newcommand{\eeqn}{\end{eqnarray}}
\newcommand{\bmath}{\begin{subequations}}
\newcommand{\emath}{\end{subequations}}
\title{Predicted electric field near small superconducting ellipsoids}
\author{J. E. Hirsch }
\address{Department of Physics, University of California, San Diego\\
La Jolla, CA 92093-0319}
 
\date{September 22, 2003} 
\begin{abstract} 
We predict the existence of large electric fields near the surface of superconducting bodies of ellipsoidal shape
of dimensions comparable to the penetration depth. The electric field is quadrupolar in nature with significant
corrections from higher order multipoles. Prolate (oblate) superconducting ellipsoids are predicted to
exhibit fields consistent with negative (positive) quadrupole moments, reflecting the fundamental charge
asymmetry of matter.
\end{abstract}
\maketitle 

Electrostatic fields cannot exist inside normal metals because the mobile electrons will move along any existing
electric field lines so as to minimize their potential energy, resulting in an equilibrium configuration with
no electric field. Any kinetic energy gained by the electron in this process will be dissipated through inelastic scattering
mechanisms.

Electrons in superconductors are also mobile (even more so than in normal metals), hence one is accustumed
to think that no electrostatic
 field can exist inside superconductors either. However, we suggest that such preconception may actually
be incorrect for the following reason: in a superconductor there are no inelastic scattering mechanisms that will dissipate the kinetic energy of 
a superfluid electron moving along an electric field line. As a consequence, as the electron moves along the field
line and decreases its potential energy it will gain kinetic energy; when it reaches the minimum potential 
energy its kinetic energy will be maximum so it will 'overshoot' and move back to a region of higher potential
energy,  leaving the electric field unscreened. Thus we argue that at least in principle the existence of electrostatic fields
inside superconductors is not ruled out by basic physics principles.

 On the contrary, we have recently proposed that an electrostatic field indeed exists inside superconductors\cite{charge,atom,efield}.
 The proposal is based on the theory of hole superconductivity\cite{hole}, that predicts that negative charge is expelled
 from the interior of the superconductor towards the surface, resulting in a net positive charge density in the
 interior, a negative charge density near the surface, and an outward pointing electric field inside the superconductor.
 
In fact, the possibility of an electrostatic field in a superconductor follows directly from London's equation for the
supercurrent $\vec{J}$ in the presence of a magnetic vector potential $\vec{A}$\cite{tinkham} 
\beq
\vec{J}=-\frac{n_s e^2}{m_e c}\vec{A}
\eeq
and Faraday's law in the form that relates  the electric field $\vec{E}$ to the electric potential $\phi$ and the magnetic vector potential: 
\beq
\vec{E}=-\vec{\nabla}\phi -\frac{1}{c} \frac{\partial\vec{A}}{\partial t} .
\eeq
Taking the time derivative of Eq. (1) and using Eq. (2) yields
\beq
\frac{\partial \vec{J}}{\partial t}=\frac{n_se^2}{m_e}(\vec{\nabla} \phi +\vec{E})
\eeq
indicating that {\it an electric field that derives from a potential will not lead to a time variation of the supercurrent}, and
in particular will not generate a supercurrent if one is not present initially.

The electric potential in the interior of the superconductor is assumed to obey the fundamental equation\cite{efield}
\beq
\phi (\vec{r})=-4\pi \lambda_L^2 \rho (\vec{r})+\phi _0(\vec{r})
\eeq
which is derived from the London equation under the assumption that the magnetic vector potential obeys the Lorenz
gauge condiction\cite{london,efield}. We postulate Eq. (4) to describe the electric potential $\phi (\vec{r})$
 and charge distribution $\rho (\vec{r})$ in the interior of
all superconductors, with $\lambda_L$ the London penetration depth and $\phi _0(\vec{r})$ the potential originating from a uniform
$positive$ charge density $\rho_0$:
\beq
\phi _0(\vec{r})=\int_V d^3r'\frac{\rho_0}{|\vec{r}-\vec{r}'|}
\eeq
where the integral is over the volume of the superconducting body. The charge density $\rho_0$ is a function of 
 parameters describing the superconducting material, of the dimensions of the body, and of temperature\cite{efield}. 
 It originates in the 'undressing' of carriers as the
system goes superconducting, as described in the theory of hole superconductivity\cite{undressing}. In particular, it increases as
the temperature is lowered below $T_c$ and the superfluid density increases, and it is larger for superconductors with
higher $T_c$.

The electric potential also obeys the usual Poisson equation
\beq
 \nabla^2\phi (\vec{r})=-4\pi\rho(\vec{r})
 \eeq
in the interior of the superconductor, and Laplace's equation in the exterior. Furthermore we assume that
$\phi (\vec{r})$ as well as its normal derivative are continuous across the surface of the
superconductor, i.e. that no surface charge density  exists.

Equation (4) predicts that $\rho(\vec{r})=\rho_0$ deep in the interior of superconductors of dimensions much larger than the
penetration depth, and an excess of negative charge within a layer of thickness $\lambda_L$ of the surface\cite{efield}.
The solution of Eqs. (4)-(6) was discussed in Ref.\cite{efield} for the case of spherical superconductors. For that
case, no electric field exists in the exterior of the superconductor and the proposed scenario cannot be detected.
Fortunately, the situation is different for superconductors of non-spherical shape.

We consider here ellipsoids of revolution, defined by 
\beq
\frac{\rho^2}{a^2}+\frac{z^2}{b^2}=1
\eeq
with $\rho,z$ cylindrical coordinates. The dipole moment of any charge distribution with the symmetry of the ellipsoid
is zero, but the quadrupole moment
\beq
Q=\int d^3r \rho(\vec{r}) [3z^2-r^2]
\eeq
is not. For the uniformly charged ellipsoid with charge density $\rho_0$ the quadrupole moment is
\beq
Q_0=\frac{8}{15} \pi a^2 b (b^2-a^2)\rho_0
\eeq
so that it is positive (negative) for prolate ($b>a$) (oblate ($b<a$)) ellipsoids for positive $\rho_0$.

We solve the differential equation resulting from Eqs. (4)-(6)
\beq
\phi (\vec{r})=\lambda_L^2 \nabla^2\phi (\vec{r})+\phi _0(\vec{r})
\eeq
numerically in the interior of the ellipsoid using the GENCOL algorithm\cite{gencol}.
Initially we assume an arbitrary boundary condition consistent with overall charge neutrality, for example that the normal derivative  
$\partial \phi/ \partial n=0$ everywhere on the surface
(Neumann boundary condition), which implies non-existence of electric fields in the exterior of the superconductor.
However  $\partial \phi /\partial n=0$ on the surface implies $\phi$ is constant on the surface. The first iteration of the
procedure yields a $\phi$ on the surface that is not constant (except for the case $b=a$), implying that the original assumption
of no electric field in the exterior was incorrect.

To achieve self-consistency we obtain the electric potential in the exterior of the superconductor using
\beq
\phi (\vec{r})=\int_V d^3r'\frac{\rho (\vec{r}')}{|\vec{r}-\vec{r}'|}
\eeq
where $\rho (\vec{r})$ is obtained from Eq. (4) using the solution obtained for $\phi(\vec{r})$ in the interior.
For the next iteration step we can use as boundary condition for the interior problem the normal derivative
$\partial \phi /\partial n$ obtained from the exterior problem or the potential itself, obtained from Eq. (11) on the
surface. After a few iterations the procedure converges to a unique solution $\phi(\vec{r})$ which is continuous
and has continuous derivatives across the surface. 

\begin{figure}
 \includegraphics[height=.4\textheight]{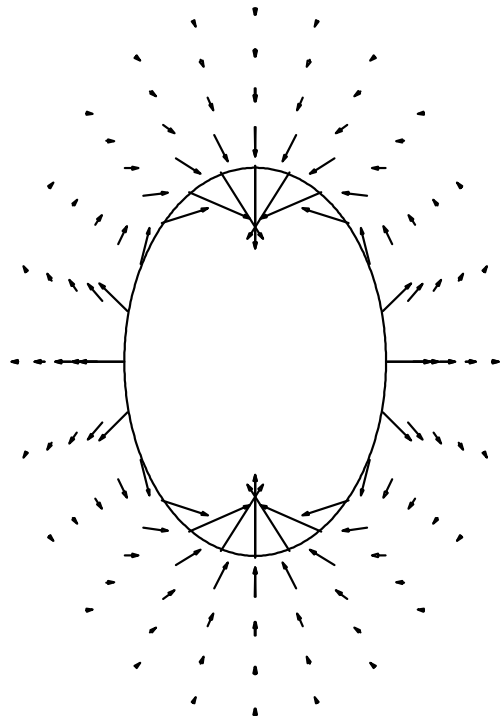}
  \caption{Electric field outside a prolate ellipsoid of dimensions $a=1$, $b=1.5$. London penetration
  depth is $\lambda_L=0.5$, and $\rho_0=1$. The arrows point in the direction of the electric field and
  their length is proportional to the magnitude of the field.}
\end{figure}

For a prolate ellipsoid with $b/a=1.5$ and London penetration depth $\lambda_L/a=0.5$ the quadrupole moment
of the resulting charge distribution is $Q=-.485$, in units so that $\rho_0=a=1$. For comparison the quadrupole
moment of the bare positive ellipsoid for this case is $Q_0=\pi$ . Figure 1 shows 
the electric field configuration outside the ellipsoid obtained (the electric field inside  is not shown  because it is
much larger). The electric field points $out$ near the surface for $\rho\sim a, z\sim 0$
and points $in$ near the surface for $\rho\sim 0, z\sim b$. For an oblate ellipsoid the situation would be reversed.
The tangential component of the field on the surface shown in Fig. 1 exerts a force on the electrons in the
direction of smaller $|z|$. As argued earlier, we interpret that 
electrons do not rearrange according to this force to screen the field because they will
increase their kinetic energy when they move in the direction of the force and decrease it when they move
opposite to it. The magnitude of the electric field decays rapidly away from the surface since it is
of quadrupolar nature. 

\begin{figure}
 \includegraphics[height=.4\textheight]{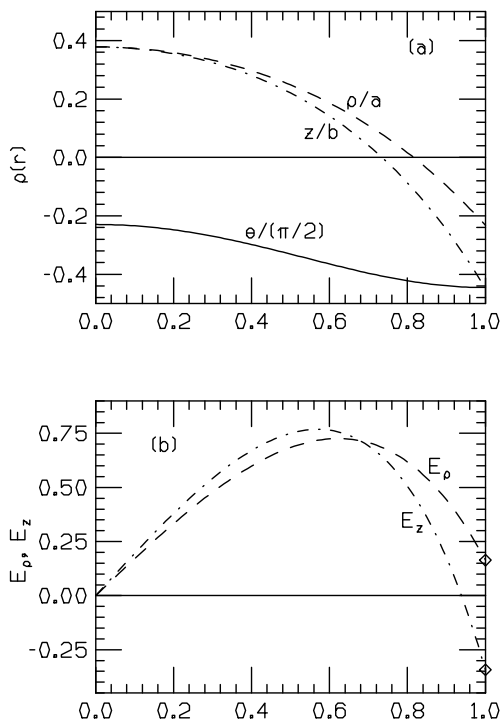}
  \caption{For the parameters of Fig. 1: (a) Charge density in the interior of the superconductor along the horizontal axis
  plotted versus $\rho / a$ (curve labeled $\rho /a $) and along the vertical axis plotted versus  $z / b$
  (curve labeled $z / b$) and on the boundary plotted versus $ \theta / (\pi / 2)$, with
  $ \theta=tan^{-1} (z / b) / (\rho / a)$); note that the negative charge density near the surface is larger in magnitude 
  along the $z$ direction. (b) Electric fields in the interior along the $\rho /a $ (dashed) and
  $z / b$ (dot-dashed) directions. Note that the electric field along the $z$ direction changes sign near the
  surface, and that the electric fields are finite at the surface.}
\end{figure}

Figure 2 shows the obtained charge density $\rho(\vec{r})$ in the interior along the axis
$z=0$ and $\rho=0$ and the electric field in the interior along these axis for the parameters
of Fig. 1. Because $\lambda_L$ is comparable to the size of the body, $\rho(r)$ does not reach
the value $\rho_0$ (=1) deep in the interior. Contrary to the case of the sphere
discussed in ref. \cite{efield}, here the charge density and electric fields are different
in the different directions. Furthermore the negative charge near the surface cannot fully screen the
electric field as in the case of the sphere, so that the field is finite at the surface of the ellipsoid
and leaks out. Note that the obtained negative charge density
near the surface is larger in the direction of the larger axis of the ellipsoid. This is found to be the case
generally for both prolate and oblate ellipsoids, which implies that the quadrupole moment will always be
negative (positive) for prolate (oblate) ellipsoids. This would even be the case if the charge density near the surface was
equal in the directions of large and small axis, but this feature enhances the effect even more.
For smaller values of the penetration depth the charge density is larger in the interior and more negative
near the surface as in the case of the sphere\cite{efield}, and a finite electric field always occurs at
the surface except for $a=b$.

\begin{figure}
 \includegraphics[height=.5\textheight]{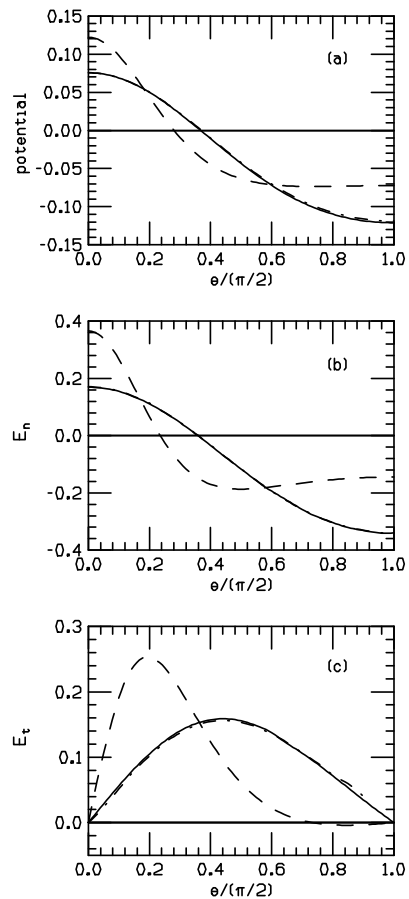}
  \caption{Electric potential (a) and electric field in the normal (b) and in the tangential (c) directions on the
  surface of the ellipsoid for the parameters of Fig. 1
  plotted versus $ \theta / (\pi / 2)$, with
  $ \theta=tan^{-1} (z / b) / (\rho / a)$). The solid and dashed-dotted lines give the results obtained
  from the inside and the outside solutions, as described in the text; the dashed lines give the corresponding potential
  and fields for a quadrupole at the origin of the same magnitude as the quadrupole moment of the
  charge distribution for this case, $Q=-0.485$.}
\end{figure}

Figure 3 shows the electric potential as well as the normal and tangential components of the electric field
on the surface of a prolate ellipsoid for the parameters of Fig. 1. The  agreement of the results from the 
inside and outside solutions indicates that the numerical iteration procedure has  converged.  The electric potential
is larger near the equator ($\theta=0$) than near the poles ($\theta=\pi/2$), consistent with a tangential electric field in the
polar direction and a normal field that points out near the equator and in near the poles. The results differ
substantially from what would be obtained from a pure quadrupole with rotational symmetry along the
$z$ axis, where the potential is given by
\beq
\phi_Q=\frac{Q}{4} \frac{3z^2-r^2}{r^5}
\eeq
These results are shown in Fig. 3 as dashed lines for comparison. This implies that there is substantial contribution
from higher order multipoles. Moving away from the surface we find that the pure quadrupolar behavior 
predicted by Eq. (12) rapidly becomes dominant.   It is easy to obtain the contribution from the different
higher order multipoles by either comparing the potential outside for different distances or by
computing integrals of the obtained charge distribution inside.  A precise experimental study of the field
distribution in the neighborhood of superconducting ellipsoids should give detailed information on the charge distribution inside the superconductor.

As $\lambda_L$ decreases compared with the dimensions of the body the electric fields increase, and
the electric field in the interior
approaches the one of a uniformly charged ellipsoid of charge density $\rho_0$, given by
$E_{\rho}=C_{\rho}\rho_0\rho$, $E_{z}=C_{z}\rho_0 z$, with $C_{\rho}, C_{z}$ scale-independent
geometrical constants.
The electric fields at the surface are a fraction (whose value depends on $b/a$)
 of the fields at the surface of the uniformly charged ellipsoid. 
 For example for $b/a=1.5$ the limiting values for the fields at the surface are
approximately $E_{\rho}^s \sim  0.13 E_{\rho}^{max}$, $E_{z}^s \sim  -0.3  E_{z}^{max}$, with
 $E_{\rho}^{max},  E_{z}^{max}$ the maximum values of the fields inside, i.e. 
$C_{\rho}\rho_0a$ and $C_{z}\rho_0 b$ respectively. The difference in potential at the surface
 between the poles and the equator reaches a limiting value which for the case $b/a=1.5$ is approximately
 $\Delta \phi \sim 1.2 E_{\rho}^s a$.
 
 As discussed in ref. \cite{efield}, energy considerations show that $\rho_0$ decreases inversely with the
 linear dimension of the sample. Consequently  the electric fields at the surface first increase as the sample
 size increases and then reach their limiting values once the sample is
 much larger than the penetration depth. The potential difference $\Delta \phi$ at the surface increases linearly
 with the linear size of the sample when the fields have reached their limiting values. We
 estimated in ref. \cite{efield} the maximum electric field in the interior of both high $T_c$ cuprates and
 $Nb$ to be of order $10^6 V/cm$, which according to the above discussion then yields for ellipsoids with $b/a\sim 1.5$
 electric fields near the surface of order
 $10^5 V/cm$ for samples larger than the penetration depth, and potential difference between polar and 
 equatorial points on the surface $\Delta \phi \sim 10^5V\times a(cm)$,
 with $a$ the linear dimension of the body.

 Such large electric fields are $not$ seen near the surface of macroscopic superconductors. We propose that
 this is so because when the potential difference $\Delta \phi$ on the surface becomes larger than the work function $W$,
 electrons will 'pop out' of the superconductor near the region of low potential  and migrate to the region of high potential on the surface, and the resulting 
 electronic layer $outside$ the superconductor will screen the electric field so that it becomes unobservable.
 For the example discussed above with a work function $W\sim 5eV$ this implies that the predicted large
 electric fields  will only be observed for samples with linear dimensions smaller than $a \sim 5000 A$. 
 We emphasize however that this estimate is very rough and depends on intrinsic parameters of the superconducting
 material\cite{efield} and on the shape parameters. The maximum sample size for which the fields are unscreened will
 increase as the temperature increases (since $\rho_0$ will decrease) and the magnitude of the fields
will decrease, such that the potential difference on the surface never exceeds the work function.

 \begin{figure}
 \includegraphics[height=.2\textheight]{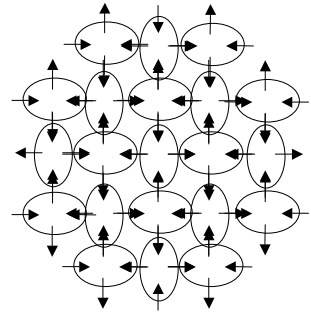}
  \caption{ Superconducting ellipsoids will tend to bunch into larger spherical shapes. The arrows indicate
  the direction of the electric fields at the surface.}
\end{figure}

 Note also  that an attractive or repulsive electric force will exist between small superconducting ellipsoids
 depending on their relative orientation. A set of small superconducting ellipsoids in close proximity will lower their energy by adopting the proper orientation
and coming close to each other, thus bunching into spherical aggregates as shown in Fig. 4. We suggest that
this may be
related to the remarkable experimental finding of formation of superconducting balls reported by Tao and
coworkers\cite{tao}.

Unlike the conventional description that is non-Lorentz invariant, Eq. (1) with $\vec{A}$ in the Lorenz gauge  leads to a relativistically covariant
description of the electrodynamics of superconductors\cite{electrodyn,london,newwork}, which we do not believe  violates any basic physical law nor contradicts any known experimental fact. 
The resulting electric field in the interior of superconductors does not lead to a time dependent supercurrent because Eq. (3) replaces
the conventional
relation between current and electric field for 'perfect conductors' usually assumed to be valid for superconductors. 
 Further experimental consequences and the relation with microscopic theory will be the subject of future work.

\acknowledgements
The author is grateful to E.N. Houstis, W.F. Mitchell, and J.R. Rice for making their GENCOL software available
for public use, and to J. Kuti for help with implementation of the software.

\end{document}